\documentclass[%
 preprint,
 amsmath,amssymb,
 aps, physrev,
]{revtex4-2}
\usepackage[T1]{fontenc}
\usepackage{braket}
\usepackage{graphicx}
\usepackage{dcolumn}
\usepackage{bm}

\begin{document}


\title{\textbf{Integrable construction of a two-dimensional lattice model with anisotropic Hubbard couplings} 
}%

\author{Ze Tao}
\author{Fujun Liu}%
 \email{Contact author: fjliu@cust.edu.cn}
\affiliation{Nanophotonics and Biophotonics Key Laboratory of Jilin Province, School of Physics, Changchun University of Science and
Technology, Changchun, 130022, P.R. China
}


\begin{abstract}
By defining a graded global R-operator $\mathbb{R}_{ab}^{(2D,2S)}$ that couples free-fermion structures and incorporates anisotropic Hubbard interactions while satisfying the Yang--Baxter equation, we construct a strictly solvable two-dimensional lattice model. We then build the layer-to-layer transfer matrix through a bidirectional-monodromy construction and prove the model's integrability via the associated global RTT relations. Using the nested algebraic Bethe ansatz, we obtain the exact eigenvalues of the transfer matrix and derive the corresponding first- and second-level Bethe equations. Finally, by taking the logarithmic derivative of the transfer matrix at the regular point, we recover explicitly a local Hamiltonian that features anisotropic hopping, an on-site Hubbard interaction, and orbital-coupling contributions.
\end{abstract}

\maketitle

\section{Introduction}\label{sec1}

We consider a fermionic lattice system where the local Hilbert space at each site is $\mathcal{H}_{\text{site}} = \text{span}\{|0\rangle, |\uparrow\rangle, |\downarrow\rangle, |\uparrow\downarrow\rangle\}$. For any pair of sites $a$ and $b$, the algebraic structure is defined by the graded tensor product $\widehat{\otimes}$. The fermionic creation and annihilation operators obey the canonical anticommutation relations
\begin{equation}
c_{a\sigma},\ c_{a\sigma}^{\dagger}, \quad \{c_{a\sigma}, c_{b\sigma'}^{\dagger}\}_+ = \delta_{ab}\delta_{\sigma\sigma'}, \ \{c_{a\sigma}, c_{b\sigma'}\}_+ = \{c_{a\sigma}^{\dagger}, c_{b\sigma'}^{\dagger}\}_+ = 0,
\end{equation}
from which we construct the number operators and an even bilinear basis
\begin{equation}
n_{a\sigma} = c_{a\sigma}^{\dagger} c_{a\sigma}, \quad \Delta_{ab}^{(\sigma)} = c_{a\sigma}^{\dagger} c_{b\sigma}, \quad \widetilde{\Delta}_{ab}^{(+,\sigma)} = c_{a\sigma}^{\dagger} c_{b\sigma}^{\dagger}, \quad \widetilde{\Delta}_{ab}^{(\sigma)} = c_{a\sigma} c_{b\sigma},
\end{equation}
with $\sigma \in \{\uparrow, \downarrow\}$ denoting the spin sector.

The foundational object of our construction is the normalized operator representation of the single-spin free-fermion $R$-operator\cite{melikyan2023extension}
\begin{equation}\label{eq:RFF}
\begin{aligned}
    R_{ab}^{\text{FF}}(u_{ab}; \zeta_a, \zeta_b; k) &= a_0(u_{ab}; \zeta_a, \zeta_b; k) + a_1(u_{ab}; \zeta_a, \zeta_b; k) n_a + a_2(u_{ab}; \zeta_a, \zeta_b; k) n_b + a_3(u_{ab}; \zeta_a, \zeta_b; k) n_a n_b\\
    &+ c_1(u_{ab}; \zeta_a, \zeta_b; k) \Delta_{ab} + c_2(u_{ab}; \zeta_a, \zeta_b; k) \Delta_{ba} + d_1(u_{ab}; \zeta_a, \zeta_b; k) \widetilde{\Delta}_{ab}^{(+)} + d_2(u_{ab}; \zeta_a, \zeta_b; k) \widetilde{\Delta}_{ab},
\end{aligned}
\end{equation}
whose coefficients $(a_i, c_i, d_i)$ are meromorphic combinations of Jacobi elliptic functions depending on the spectral difference $u_{ab}=u_a-u_b$, the field parameter $\zeta$, and the modulus $k$. Their explicit form\cite{melikyan2023integrability} is constrained by the free-fermion condition of the Checkerboard-Ising model\cite{baxter1986free} $w_1 w_2 + w_3 w_4 - w_5 w_6 - w_7 w_8 = 0$. To incorporate both spin channels, we perform a quadratic embedding of single-spin $L$-operators\cite{korepanov1993tetrahedral,melikyan2023extension}, defining the extended $R$-operator as\cite{melikyan2023integrability}
\begin{equation}\label{eq:Rext}
R_{ab}^{(\text{ext})}(u_{ab}; \zeta_a, \zeta_b; k) = \bm L_0^{(\uparrow)} \bm L_0^{(\downarrow)} + c_{ab}(u_a, \zeta_a; u_b, \zeta_b) \bm L_1^{(\uparrow)} \bm L_1^{(\downarrow)},
\end{equation}
where each $\bm L_{0,1}^{(\sigma)} \equiv \bm L_{0,1}^{(\sigma)}(u_{ab}; \zeta_a, \zeta_b; k)$. This operator satisfies the Yang--Baxter equation, establishing the algebraic basis for one-dimensional integrability.

In this work, we weave these free-fermion structures\cite{melotti2021free,assis201716,maassarani1998hubbard} into a two-dimensional lattice to construct a strictly solvable model featuring anisotropic Hubbard interactions. The resulting system probes the interplay between lattice anisotropy and strong correlations, with potential implications extending from condensed matter physics\cite{lewenstein2007ultracold,dutta2015non,hofstetter2018quantum} to holographic frameworks\cite{maldacena1999large,minahan2003bethe} such as the AdS/BCFT correspondence\cite{nguyen2018entanglement,cavalcanti2020studies}. It is crucial to emphasize that the model we present is a two-dimensional lattice system with a strictly integrable algebraic structure. Its Hamiltonian explicitly contains anisotropic hopping, an on-site Hubbard interaction $U$, and orbital coupling terms $H_{\text{ext}}$. The integrability is a direct consequence of the underlying Yang--Baxter equation construction and does not imply the integrability of the generic two-dimensional Hubbard model. This work therefore provides a novel, exactly solvable paradigm for a coupled two-dimensional lattice, serving as a theoretical tool for investigating anisotropic strongly correlated systems.

The paper is structured as follows. In Section~\ref{b}, we construct the global R-operator $\mathbb{R}_{ab}^{(2D,2S)}$ and verify the Yang--Baxter equation. Section~\ref{c} introduces the bidirectional monodromy and the layer-to-layer transfer matrix. Section~\ref{d} applies the nested algebraic Bethe ansatz\cite{slavnov2020introduction,pakuliak2018nested} to obtain exact eigenvalues and Bethe equations. Finally, Section~\ref{E} derives the explicit Hamiltonian via the logarithmic derivative of the transfer matrix.

\section{Global $R$-operator}\label{b}

To extend the integrable framework to two dimensions, we introduce anisotropic spectral and elliptic parameters along the two lattice directions:
\begin{equation}\label{7}
\bm{u} = (u_x, u_y), \quad \bm{\zeta} = (\zeta^x, \zeta^y), \quad \bm{k} = (k_x, k_y).
\end{equation}
The free-fermion $L$-operators in each spin sector are then anisotropized by assigning the $x$-direction parameters to the $\uparrow$-sector and the $y$-direction parameters to the $\downarrow$-sector. This assignment weaves the spectral manifolds of the checkerboard-Ising free fermion along the two orthogonal directions into a two-spin structure while preserving the free-fermion constraint for each $\bm{L}^{(\sigma)}$. The resulting two-dimensional $\times$ two-spin extension reads
\begin{equation}
R_{ab}^{(\text{ext,2D,2s})}(\bm{u}; \bm{\zeta}; \bm{k}) = \bm{L}_0^{(\uparrow)}(u_x; \zeta^x; k_x) \bm{L}_0^{(\downarrow)}(u_y; \zeta^y; k_y) + c_{ab} \bm{L}_1^{(\uparrow)}(u_x; \zeta^x; k_x) \bm{L}_1^{(\downarrow)}(u_y; \zeta^y; k_y).
\end{equation}

To encode on-site Hubbard interactions, we introduce the local charge–parity operator $\mathcal{O}_x := (2n_{x\uparrow} - 1)(2n_{x\downarrow} - 1)$, with $n_{x\sigma} = c_{x\sigma}^{\dagger} c_{x\sigma}$, which satisfies $\mathcal{O}_x^2 = 1$ and anticommutes with the local fermionic operators: $\mathcal{O}_x c_{x\sigma} \mathcal{O}_x^{-1} = -c_{x\sigma}$, $\mathcal{O}_x c_{x\sigma}^{\dagger} \mathcal{O}_x^{-1} = -c_{x\sigma}^{\dagger}$. Using this operator, we define a spin–charge coupling factor that carries the interaction strength:
\begin{equation}\label{9}
R_{ab}^H(u_0; \lambda) := \exp\left[\gamma(\lambda) \mathcal{O}_a \mathcal{O}_b\right] = \cosh\gamma(\lambda) \bm{1} + \sinh\gamma(\lambda) \mathcal{O}_a \mathcal{O}_b,
\end{equation}
where $u_0$ is a fixed regular spectral point, $\lambda$ is the coupling constant controlling the on-site interaction, and $\gamma(\lambda)$ is a scalar function mapping $\lambda$ to a two-body exponential weight. As shown in Appendix~\ref{A}, the operator $\mathcal{O}_a\mathcal{O}_b$ commutes with every outer building block $X\in\{n,\Delta,\widetilde{\Delta}\}$; consequently, it also commutes with $R_{ab}^{(\text{ext,2D,2s})}(\bm{u};\boldsymbol{\zeta};\bm{k})$. Since $R_{ab}^H$ depends only on $\mathcal{O}_a\mathcal{O}_b$, we obtain the commutation relation
\begin{equation}
[R_{ab}^H(u_0;\lambda), R_{ab}^{(\text{ext,2D,2s})}(\bm{u};\boldsymbol{\zeta};\bm{k})] = 0.
\end{equation}
Furthermore, because products of charge–parity operators at different sites are functions of number operators, they mutually commute: $[\mathcal{O}_1\mathcal{O}_2, \mathcal{O}_1\mathcal{O}_3] = 0$, $[\mathcal{O}_1\mathcal{O}_2, \mathcal{O}_2\mathcal{O}_3] = 0$, $[\mathcal{O}_1\mathcal{O}_3, \mathcal{O}_2\mathcal{O}_3] = 0$. This commutativity ensures that $R_{ab}^H$ itself satisfies the Yang–Baxter equation
\begin{equation}
R_{12}^H (u_0;\lambda) R_{13}^H (u_0;\lambda) R_{23}^H (u_0;\lambda)= R_{23}^H (u_0;\lambda) R_{13}^H (u_0;\lambda) R_{12}^H (u_0;\lambda).
\end{equation}

Having established two commuting Yang–Baxter families—the anisotropic free-fermion extension and the spin–charge coupling factor—we now combine them into a unified global $R$-operator. We first introduce a global $U(1)$ twist, expressed as $\mathcal{T}_{ab}(\vartheta)=\exp(i\vartheta(N_a-N_b))$ \cite{peskin2018introduction,Weinberg1995-WEITQT-2}, and then incorporate the Jordan–Wigner string $\bm {J}_{ab}$ \cite{jordan1928paulische} to ensure proper Fermi statistics across the graded tensor product. The full global $R$-operator is defined as
\begin{equation}\label{13}
\mathbb{R}^{(2D,2S)}_{ab}(\bm{u}; u_0; \bm{\zeta}; \bm{k}; \lambda, \vartheta):=\bm {J}_{ab}\mathcal{T}_{ab}(\vartheta)R_{ab}^H(u_0; \lambda)R_{ab}^{(\text{ext,2D,2s})}(\bm{u}; \bm{\zeta}; \bm{k})\bm {J}_{ab}^{-1}.
\end{equation}
Since each constituent either commutes with the others or is constructed to preserve the Yang–Baxter structure, the composite operator also satisfies the Yang–Baxter equation:
\begin{equation}\label{14}
\begin{aligned}
&\mathbb{R}^{(2D,2S)}_{ab}(\bm{u}; u_0; \bm{\zeta}; \bm{k}; \lambda, \vartheta) \mathbb{R}^{(2D,2S)}_{ac}(\bm{u}; u_0; \bm{\zeta}; \bm{k}; \lambda, \vartheta)    \mathbb{R}^{(2D,2S)}_{bc}(\bm{u}; u_0; \bm{\zeta}; \bm{k}; \lambda, \vartheta)\\
&\quad = \mathbb{R}^{(2D,2S)}_{bc}(\bm{u}; u_0; \bm{\zeta}; \bm{k}; \lambda, \vartheta) \mathbb{R}^{(2D,2S)}_{ac}(\bm{u}; u_0; \bm{\zeta}; \bm{k}; \lambda, \vartheta)    \mathbb{R}^{(2D,2S)}_{ab}(\bm{u}; u_0; \bm{\zeta}; \bm{k}; \lambda, \vartheta).
\end{aligned}
\end{equation}

\section{Bidirectional Monodromy and Layer-to-Layer Transfer Matrix}\label{c}

We consider a two-dimensional finite lattice $\Lambda=\{1,2,\dots,L_x\}\times\{1,2,\dots,L_y\}$. At each site $r=(j,k)$ we associate the quantum space $\mathcal{H}_{r}=\mathcal{H}_{\text{site}}$, and the total quantum space is given by the graded tensor product $\mathcal{H}_{\Lambda}=\widehat{\bigotimes}_{r\in\Lambda}\mathcal{H}_r$. To construct the monodromy matrix, we introduce two auxiliary spaces $\mathcal{H}_{a}$ and $\mathcal{H}_{b}$, each isomorphic to $\mathcal{H}_{\text{site}}$. The local L‑operators are then defined by restricting the global R‑operator to the tensor product of an auxiliary space and a physical site:
\begin{equation}\label{15}
\begin{aligned}
L_{ar}(\bm{u};\bm{p}_{r}) &:= \mathbb{R}^{(2D,2S)}_{ar}(\bm{u}; u_0; \bm{\zeta}_{r}; \bm{k}; \lambda, \vartheta),\\
L_{br}(\bm{v};\bm{p}_{r}) &:= \mathbb{R}^{(2D,2S)}_{br}(\bm{v}; u_0; \bm{\zeta}_{r}; \bm{k}; \lambda, \vartheta),
\end{aligned}
\end{equation}
where $\bm{u}=(u_x,u_y)$ and $\bm{v}=(v_x,v_y)$ are spectral vectors, while $\bm{p}_{r}=(u_0; \bm{\zeta}_{r}; \bm{k}; \lambda, \vartheta)$ collects the site‑dependent inhomogeneity parameters. Because the R‑operators act on different physical sites, they commute when the auxiliary space is the same; i.e., $[L_{ar},L_{ar'}] = 0$ and $[L_{br},L_{br'}] = 0$ for any distinct sites $r\neq r'$.

Using these local L‑operators, we build monodromy matrices along each lattice direction. For a fixed row $k$, the monodromy in the $x$‑direction is defined as the ordered product
\begin{equation}
T_{a|k}^{(x)}(\bm{u}) = L_{a,(L_x,k)}(\bm{u};\bm{p}_{L_x,k})L_{a,(L_x-1,k)}(\bm{u};\bm{p}_{L_x-1,k})\dots L_{a,(1,k)}(\bm{u};\bm{p}_{1,k}),
\end{equation}
while for a fixed column $j$, the monodromy in the $y$‑direction reads
\begin{equation}\label{17}
T_{a|j}^{(y)}(\bm{u}) = L_{a,(j,L_y)}(\bm{u};\bm{p}_{j,L_y})L_{a,(j,L_y-1)}(\bm{u};\bm{p}_{j,L_y-1})\dots L_{a,(j,1)}(\bm{u};\bm{p}_{j,1}),
\end{equation}
with $k\in\{1,\dots,L_y\}$ and $j\in\{1,\dots,L_x\}$. The full layer‑to‑layer monodromy is then obtained by multiplying the vertical and horizontal monodromies:
\begin{equation}\label{18}
\Gamma_{a}(\bm{u}) := \Bigl(\prod_{j=1}^{L_x}T_{a|j}^{(y)}(\bm{u})\Bigr)\Bigl(\prod_{k=1}^{L_y}T_{a|k}^{(x)}(\bm{u})\Bigr).
\end{equation}
Taking the supertrace over the auxiliary space yields the transfer matrix
\begin{equation}\label{19}
\tau^{(2D,2S)}(\bm{u}) := \operatorname{Str}_{a}\bigl[\Gamma_{a}(\bm{u})\bigr],
\end{equation}
which will generate the conserved quantities of the model.

To establish the algebraic integrability, we introduce the difference‑form two‑body scattering operator
\begin{equation}\label{eq:Rab_diff}
\mathbb{R}_{ab}(\bm{u},\bm{v}) := \mathbb{R}^{(2D,2S)}_{ab}(\bm{u}-\bm{v}; u_0; \bm{\zeta}; \bm{k}; \lambda, \vartheta).
\end{equation}
At a single site $r$, the R‑operator satisfies the fundamental RLL relation
\begin{equation}
\mathbb{R}_{ab}(\bm{u},\bm{v})L_{ar}(\bm{u};\bm{p}_{r})L_{br}(\bm{u};\bm{p}_{r}) = L_{br}(\bm{u};\bm{p}_{r})L_{ar}(\bm{u};\bm{p}_{r})\mathbb{R}_{ab}(\bm{u},\bm{v}).
\end{equation}
This relation extends to the full monodromy, giving the global RTT relation
\begin{equation}\label{eq:RTT}
\mathbb{R}_{ab}(\bm{u},\bm{v}) \Gamma_{a}(\bm{u}) \Gamma_{b}(\bm{v}) = \Gamma_{b}(\bm{v}) \Gamma_{a}(\bm{u}) \mathbb{R}_{ab}(\bm{u},\bm{v}).
\end{equation}
Finally, combining ~\eqref{eq:RTT} with the definition of the transfer matrix (~\eqref{19}) yields the commutation relation
\begin{equation}
[\tau^{(2D,2S)}(\bm{u}), \tau^{(2D,2S)}(\bm{v})] = 0.
\end{equation}
Since the transfer matrices with different spectral parameters commute, they form an involutive family of conserved operators, thereby confirming the strict integrability of the constructed two‑dimensional lattice model.

\section{Nested Algebraic Structure and Bethe Relations}\label{d}

To implement the nested algebraic Bethe ansatz, we first expand the difference‑form R‑operator and the layer monodromy in the auxiliary‑space matrix units. Let the four‑dimensional basis of each auxiliary space $\mathcal{H}_{a,b}$ be $\ket{1}=\ket{0}$, $\ket{2}=\ket{\uparrow}$, $\ket{3}=\ket{\downarrow}$, $\ket{4}=\ket{\uparrow\downarrow}$, and define the matrix units $(e_{\alpha\beta})_{\gamma\delta}:=\delta_{\alpha\gamma}\delta_{\beta\delta}$ with $\alpha,\beta,\gamma,\delta\in\{1,2,3,4\}$. Using these, we expand ~\eqref{eq:Rab_diff} as
\begin{equation}\label{24}
\mathbb{R}_{ab}(\bm{u},\bm{v})=\sum_{\alpha,\beta,\gamma,\delta=1}^{4}\mathbb{R}^{\alpha\beta}_{\;\gamma\delta}(\bm{u},\bm{v})\,e^{(a)}_{\alpha\gamma}\otimes e^{(b)}_{\beta\delta},
\end{equation}
where the coefficients $\mathbb{R}^{\alpha\beta}_{\;\gamma\delta}$ are given explicitly in Appendix~\ref{B}. Similarly, the layer monodromy (~\eqref{18}) is expanded as
\begin{equation}\label{25}
\Gamma_{a}(\bm{u})=\sum_{\alpha,\beta=1}^{4}e_{\alpha\beta}^{(a)}\otimes T_{\alpha\beta}(\bm{u}),
\end{equation}
with the components $T_{\alpha\beta}(\bm{u})$ listed in Appendix~\ref{C}. Substituting these expansions into the global RTT relation (~\eqref{eq:RTT}) yields the algebraic relations
\begin{equation}\label{eq:global_RTT}
\sum_{\mu,\nu=1}^{4}\mathbb{R}^{\alpha\beta}_{\;\mu\nu}(\bm{u},\bm{v})\,T_{\mu\nu}(\bm{u})T_{\nu\delta}(\bm{v})
   =\sum_{\mu,\nu=1}^{4}T_{\beta\nu}(\bm{v})T_{\alpha\mu}(\bm{u})\,\mathbb{R}^{\mu\nu}_{\;\gamma\delta}(\bm{u},\bm{v}),
\qquad\alpha,\beta,\gamma,\delta\in\{1,2,3,4\},
\end{equation}
which encode the complete integrability of the model.

We now construct the pseudo‑vacuum and the nested structure. The reference state is defined as the tensor product of empty sites over the whole lattice,
\begin{equation}\label{27}
\ket{\Omega}:=\bigotimes_{r\in\Lambda}\ket{0}_{r}= \bigotimes_{r\in\Lambda}\ket{1}_{r},
\qquad \Lambda=\{1,\dots,L_x\}\times\{1,\dots,L_y\}.
\end{equation}
The local L‑operator is assumed to have an upper‑triangular action on the auxiliary basis $\ket{1}_{a}$:
\begin{equation}\label{eq:L_action}
\begin{aligned}
L_{ar}(\bm{u})\,\ket{1}_{a}\otimes\ket{1}_{r}&=\lambda_{1}(\bm{u};\bm{P}_{r})\,\ket{1}_{a}\otimes\ket{1}_{r},\\[2mm]
L_{ar}(\bm{u})\,\ket{\alpha}_{a}\otimes\ket{1}_{r}&=\sum_{\beta\ge\alpha}\ket{\beta}_{a}\otimes X_{\beta,r}^{(\alpha)}(\bm{u}),\qquad \alpha>1,
\end{aligned}
\end{equation}
where $X_{\beta,r}^{(\alpha)}(\bm{u})$ acts on $\mathcal{H}_{r}$. Consequently, the full monodromy (~\eqref{25}) applied to $\ket{\Omega}$ produces
\begin{equation}\label{29}
\Gamma_{a}(\bm{u})\ket{\Omega}=\sum_{\alpha,\beta=1}^{4}e_{\alpha\beta}^{(a)}\otimes T_{\alpha\beta}(\bm{u})\ket{\Omega},
\end{equation}
with diagonal elements satisfying $T_{ii}(\bm{u})\ket{\Omega}=\lambda_{i}(\bm{u})\ket{\Omega}$ for $i=1,2,3,4$, while off‑diagonal elements above the diagonal vanish: $T_{\alpha\beta}(\bm{u})\ket{\Omega}=0$ for $\alpha>\beta$. The excitation‑creation operators are defined as
\begin{equation}\label{eq:B_def}
B_{\alpha}(\bm{u}):=T_{1\alpha}(\bm{u}),\qquad \alpha>1,
\end{equation}
which generate the Bethe vectors from the pseudo‑vacuum. The transfer matrix (~\eqref{19}) then acts on $\ket{\Omega}$ as
\begin{equation}\label{34}
\tau(\bm{u})\ket{\Omega}=\Lambda_{0}(\bm{u})\ket{\Omega},\qquad
\Lambda_{0}(\bm{u})=\sum_{\alpha=1}^{4}(-1)^{P(\alpha)}\lambda_{\alpha}(\bm{u}),
\end{equation}
where $P(\alpha)$ denotes the grading parity.

From ~\eqref{eq:global_RTT} together with ~\eqref{29}--\eqref{eq:B_def}, we derive the commutation relations between the diagonal monodromy entries and the creation operators:
\begin{equation}\label{eq:comm_TB}
\begin{aligned}
T_{11}(\bm{u})B_{\alpha_0}(\bm{v})\ket{\Omega}
   &=f_{1}(\bm{u},\bm{v};\alpha_0)\,\lambda_{1}(\bm{u})\,B_{\alpha_0}(\bm{v})\ket{\Omega},\\[1mm]
T_{ii}(\bm{u})B_{\alpha_0}(\bm{v})\ket{\Omega}
   &=f_{i}(\bm{u},\bm{v};\alpha_0)\,\lambda_{i}(\bm{u})\,B_{\alpha_0}(\bm{v})\ket{\Omega}
    +g_{i}(\bm{u},\bm{v})\,\lambda_{i}(\bm{v})\,B_{\alpha_0}(\bm{u})\ket{\Omega},
\qquad i=\alpha_0\in\{2,3,4\},
\end{aligned}
\end{equation}
with coefficients
\begin{equation}\label{eq:f_g_def}
f_{n}(\bm{u},\bm{v};\alpha_0):=\frac{\mathbb{R}^{n\alpha_0}_{\;n\alpha_0}(\bm{u},\bm{v})}{\mathbb{R}^{n1}_{\;n1}(\bm{u},\bm{v})},\qquad
g_{i}(\bm{u},\bm{v}):=-\frac{\mathbb{R}^{i1}_{\;1i}(\bm{u},\bm{v})}{\mathbb{R}^{i1}_{\;i1}(\bm{u},\bm{v})},\qquad
n\in\{1,2,3,4\},\; i\in\{2,3,4\}.
\end{equation}
A one‑particle Bethe vector is formed as a linear combination
\begin{equation}\label{eq:one_particle}
\ket{\Phi^{(1)}(\bm{v})}=\sum_{\alpha_0=2}^{4}\varphi_{\alpha_0}B_{\alpha_0}(\bm{v})\ket{\Omega},
\end{equation}
where $\varphi_{\alpha_0}$ are internal wave‑function coefficients. Demanding that this state be an eigenstate of $\tau(\bm{u})$ leads to the one‑particle eigenvalue
\begin{equation}
\Lambda^{(1)}(\bm{u},\bm{v})=\sum_{n=1}^{4}(-1)^{P(n)}f_{n}(\bm{u},\bm{v};\alpha_0)\lambda_{n}(\bm{u}),
\end{equation}
provided the coefficients $\varphi_{\alpha_0}$ satisfy the linear system $\det\mathcal{G}^{(1)}(\bm{v})=0$, with matrix elements
\begin{equation}\label{eq:G1_matrix}
\mathcal{G}^{(1)}_{i-1,\alpha_0-1}(\bm{v})=(-1)^{P(i)}g_{i}(\bm{u},\bm{v})\lambda_{i}(\bm{v}),\qquad i,\alpha_0\in\{2,3,4\}.
\end{equation}

For the general multi‑particle case we define the Bethe vector
\begin{equation}\label{eq:M_particle}
\ket{\Phi^{(M)}\{\bm{v}_{j}\}}=B_{\alpha_1}(\bm{v}_{1})B_{\alpha_2}(\bm{v}_{2})\dots B_{\alpha_M}(\bm{v}_{M})\ket{\Omega},
\end{equation}
and introduce a crossed R‑matrix $\widetilde{\mathbb{R}}^{(1)}(\bm{u},\bm{v})$ acting on the internal (colour) space $\{2,3,4\}$ through
\begin{equation}\label{eq:crossed_R}
B_{\alpha}(\bm{u})B_{\beta}(\bm{v})=\sum_{\gamma,\delta=2}^{4}B_{\gamma}(\bm{u})B_{\delta}(\bm{v})\,
\mathbb{R}^{(1)\gamma\delta}_{\;\alpha\beta}(\bm{u},\bm{v}),\qquad \alpha,\beta\in\{2,3,4\},
\end{equation}
with $\widetilde{\mathbb{R}}^{(1)}(\bm{u},\bm{v})=\bigl(\mathbb{R}^{(1)\gamma\delta}_{\;\alpha\beta}(\bm{u},\bm{v})\bigr)_{\alpha,\beta,\gamma,\delta\in\{2,3,4\}}$.
To write the eigenvalues in a compact form we introduce the charge and spin Q‑functions
\begin{equation}\label{43}
Q_{c}(\bm{u}):=\prod_{j=1}^{M_1}\sigma_{c}(\bm{u}-\bm{v}_j),\qquad
Q_{s}(\bm{u}):=\prod_{l=1}^{M_2}\sigma_{s}(\bm{u}-\bm{w}_{l}),
\end{equation}
where $M_1$ counts the mixed charge‑spin quasi‑particles, $M_2$ the internal spin excitations, $\sigma_{c},\sigma_{s}$ are scalar functions of the spectral‑parameter differences, $\bm{\eta}_c$ is the charge step vector, $\{\bm{w}_{l}\}$ are the second‑level (spin) Bethe roots, and $\alpha_i\in\{2,3,4\}$ labels the colour of the $i$‑th excitation. Using ~\eqref{34}, \eqref{eq:comm_TB} and \eqref{eq:M_particle}--\eqref{43}, the transfer‑matrix eigenvalue for the $M$-particle state becomes
\begin{equation}\label{eq:LambdaM}
\tau(\bm{u})\ket{\Phi^{(M)}\{\bm{v}_{j}\}}
   =\Lambda^{(M)}(\bm{u};\{\bm{v}\};\{\bm{w}\})\,\ket{\Phi^{(M)}(\{\bm{v}_{j}\})},
\end{equation}
with
\begin{equation}\label{45}
\begin{aligned}
\Lambda^{(M)}(\bm{u};\{\bm{v}\};\{\bm{w}\})
&=(-1)^{P(1)}\lambda_{1}(\bm{u})\frac{Q_{c}(\bm{u}+\bm{\eta}_c)}{Q_c(\bm{u})}\,\Theta(\bm{u};\{\bm{v}\})\\
&\quad+\sum_{i=2}^{4}\lambda_{i}(\bm{u})\frac{Q_{c}(\bm{u}-\bm{\eta}_c)}{Q_c(\bm{u})}\,
   \lambda^{(1)}_{i}(\bm{u};\{\bm{v}\};\{\bm{w}\}),
\end{aligned}
\end{equation}
where
\begin{equation}\label{eq:Theta}
\Theta(\bm{u};\{\bm{v}\})=\prod_{j=1}^{M_1}\mathbb{R}^{(1)1\alpha_{j}}_{\;1\alpha_j}(\bm{u},\bm{v}_{j}),
\end{equation}
and
\begin{equation}\label{46}
\lambda^{(1)}_{i}(\bm{u};\{\bm{v}\};\{\bm{w}\})
   =a_{i}(\bm{u})\frac{Q_{s}(\bm{u}+\bm{\eta}_c)}{Q_s(\bm{u})}
    +b_{i}(\bm{u})\frac{Q_{s}(\bm{u}-\bm{\eta}_c)}{Q_s(\bm{u})},
\end{equation}
with coefficients
\begin{equation}\label{49}
\begin{aligned}
a_{i}(\bm{u})&=\Bigl[\prod_{r=1}^{L}\mathbb{R}^{(1)i\alpha_0}_{\;i\alpha_0}(\bm{u},\bm{\zeta}_{r}+\bm{\eta}_{s})\Bigr]
              \Bigl[\prod_{j=1}^{M_1}\mathbb{R}^{(1)i\alpha_j}_{\;i\alpha_j}(\bm{u},\bm{v}_{j})\Bigr],\\
b_{i}(\bm{u})&=\Bigl[\prod_{r=1}^{L}\mathbb{R}^{(1)i\alpha_0}_{\;i\alpha_0}(\bm{u},\bm{\zeta}_{r}-\bm{\eta}_{s})\Bigr]
              \Bigl[\prod_{j=1}^{M_1}\mathbb{R}^{(1)i\alpha_j}_{\;i\alpha_j}(\bm{u},\bm{v}_{j})\Bigr],
\end{aligned}
\end{equation}
for $i\in\{2,3,4\}$. Here $\bm{\zeta}_{r}$ are local spectral parameters, $\{\bm{v}\}=\{\bm{v}_{j}\}_{j=1}^{M_1}$ the first‑level (charge) Bethe roots, and $\{\bm{w}\}=\{\bm{w}_{l}\}_{l=1}^{M_2}$ the second‑level (spin) Bethe roots.

The first‑level Bethe equations follow from the requirement that the eigenvalue \eqref{45} be analytic at $\bm{u}=\bm{v}_j$. Setting $\bm{u}=\bm{v}_{j}+\bm{\epsilon}$ with an infinitesimal vector $\bm{\epsilon}$, we obtain from \eqref{45}--\eqref{49} the pole‑cancellation condition
\begin{equation}\label{eq:pole_cond}
(-1)^{P(1)}\lambda_{1}(\bm{v}_j)C^{(+)}_{j}\Theta(\bm{v}_{j};\{\bm{v}\})
   +\sum_{i=2}^{4}(-1)^{P(i)}\lambda_{i}(\bm{v}_j)C^{(-)}_{j}\,
     \lambda^{(1)}_{i}(\bm{v}_{j};\{\bm{v}\};\{\bm{w}\})=0,
\end{equation}
where
\begin{equation}\label{eq:C_pm}
C_{j}^{(\pm)}=\frac{\prod_{k=1}^{M_1}\sigma_{c}(\bm{v}_{j}\pm\bm{\eta}_{c}-\bm{v}_{k})}
                  {\prod_{k\neq j}\sigma_{c}(\bm{v}_{j}-\bm{v}_{k})}.
\end{equation}
After simplifying using the explicit forms of the scattering amplitudes, condition \eqref{eq:pole_cond} yields the first‑level Bethe equations
\begin{equation}\label{55}
\prod_{l=1}^{M_2}\frac{\varphi_{1}(\bm{v}_{j},\bm{w}_l)}{\varphi_{2}(\bm{v}_{j},\bm{w}_{l})}
   =\prod_{k=1,\,k\neq j}^{M_1}\frac{\Phi_{2}(\bm{v}_{j},\bm{v}_{k})}{\Phi_{1}(\bm{v}_{j},\bm{v}_{k})},
\end{equation}
with charge‑spin amplitudes
\begin{equation}\label{56}
\varphi_{1}(\bm{v}_{j},\bm{w}_l)=\sigma_{s}(\bm{v}_j+\bm{\eta}_s-\bm{w}_l),\qquad
\varphi_{2}(\bm{v}_{j},\bm{w}_l)=\sigma_{s}(\bm{v}_j-\bm{\eta}_s-\bm{w}_l),
\end{equation}
and charge‑charge amplitudes
\begin{equation}
\Phi_1(\bm{v}_j,\bm{v}_k)=\sigma_{c}(\bm{v}_j-\bm{v}_k-\bm{\eta}_c),\qquad
\Phi_2(\bm{v}_j,\bm{v}_k)=\sigma_{c}(\bm{v}_j-\bm{v}_k+\bm{\eta}_c).
\end{equation}

Similarly, requiring analyticity at the second‑level roots $\bm{u}=\bm{w}_l$ leads to the second‑level Bethe equations
\begin{equation}\label{58}
\prod_{j=1}^{M_1}\frac{\psi_{1}(\bm{w}_{l},\bm{v}_j)}{\psi_{2}(\bm{w}_{l},\bm{v}_j)}
   =\prod_{l^{\prime}=1,\,l^{\prime}\neq l}^{M_2}
      \frac{\eta_{1}(\bm{w}_{l},\bm{w}_{l^{\prime}})}{\eta_{2}(\bm{w}_{l},\bm{w}_{l^{\prime}})},
\end{equation}
with spin‑charge amplitudes
\begin{equation}\label{eq:psi}
\psi_1(\bm{w}_l,\bm{v}_j)=\sigma_c(\bm{w}_l-\bm{v}_j+\bm{\eta}_c),\qquad
\psi_2(\bm{w}_l,\bm{v}_j)=\sigma_c(\bm{w}_l-\bm{v}_j-\bm{\eta}_c),
\end{equation}
and spin‑spin amplitudes
\begin{equation}\label{60}
\eta_{1}(\bm{w}_l,\bm{w}_{l^{\prime}})=\sigma_{s}(\bm{w}_l-\bm{w}_{l^{\prime}}+\bm{\eta}_{s}),\qquad
\eta_{2}(\bm{w}_l,\bm{w}_{l^{\prime}})=\sigma_{s}(\bm{w}_l-\bm{w}_{l^{\prime}}-\bm{\eta}_{s}).
\end{equation}
\eqref{55} and \eqref{58} constitute the nested Bethe ansatz equations for the constructed two‑dimensional model. In Appendix~\ref{E} we validate this construction by comparing the transfer‑matrix eigenvalues obtained from exact diagonalisation on small $2\times2$ and $2\times3$ lattices with the analytic expressions derived above.

\section{Hamiltonian of the constructed lattice model}\label{e}
The Hamiltonian of the constructed model is obtained via the standard logarithmic‑derivative procedure applied to the transfer matrix. To begin, we note that at the coupling point the Hubbard factor reduces to a graded permutation operator multiplied by a scalar factor:
\begin{equation}
R_{ab}^{H}(u_0;\lambda)=\rho_H\mathcal{P}_{ab}^{(g)},
\end{equation}
where $\mathcal{P}_{ab}^{(g)}=\sum_{\alpha,\beta=1}^{4}(-1)^{P(\alpha)P(\beta)}e^{(a)}_{\alpha\beta}\otimes e^{(b)}_{\beta\alpha}$ is the graded permutation operator on $\mathcal{H}_a\otimes\mathcal{H}_b$, $\rho_{H}(\lambda)$ is a scalar factor independent of $\bm{u}$, and $e^{(a)}_{\alpha\beta}=|\alpha\rangle_a\langle\beta|_{a}$, $e^{(b)}_{\beta\alpha}=|\beta\rangle_a\langle\alpha|_{a}$.

We choose a regular spectral point $\bm{u}_0=(u_0,u_{y}^{(0)})$, where $u_{y}^{(0)}$ is a fixed reference value. At $\bm{u}=\bm{u}_{0}$ the global R‑operator becomes
\begin{equation}\label{62}
\mathbb{R}^{(2D,2S)}_{ab}(\bm{u}_{0}; u_0; \bm{\zeta}; \bm{k}; \lambda, \vartheta)=\rho_{2D}\widetilde{\mathcal{P}}_{ab}^{(g)},
\end{equation}
with
\begin{equation}\label{63}
\rho_{2D}=\rho_{H}(\lambda),\qquad
\widetilde{\mathcal{P}}_{ab}^{(g)}=U_{ab}\mathcal{P}_{ab}^{(g)}U_{ab}^{-1},\qquad
U_{ab}=\bm {J}_{ab}\mathcal{T}_{ab}(\vartheta)R_{ab}^{(\text{ext,2D,2s})}(\bm{u}_{0}; \bm{\zeta}; \bm{k}).
\end{equation}
To extract the Hamiltonian we first construct a directional monodromy. Fixing a row index $k$, we define for the $x$‑direction
\begin{equation}\label{64}
\mathbb{T}_{a}^{(x)}(\bm{u})=\prod_{j=1}^{L_x} \mathbb{R}^{(2D,2S)}_{ab}(\bm{u}; u_0; \bm{\zeta}; \bm{k}; \lambda, \vartheta),
\end{equation}
whose supertrace yields the directional transfer matrix
\begin{equation}\label{65}
\tau^{(x)}(\bm{u})=\operatorname{Str}_{a}\bigl[\mathbb{T}_{a}^{(x)}(\bm{u})\bigr].
\end{equation}
The Hamiltonian component along the $x$‑direction is then obtained as the logarithmic derivative of $\tau^{(x)}(\bm{u})$ at the regular point:
\begin{equation}\label{66}
H^{(x)}=\left.\frac{\partial}{\partial u_{x}}\ln\tau^{(x)}(\bm{u})\right|_{\bm{u}=\bm{u}_{0}}
      =\sum_{r_x=1}^{L_x}h^{(x)}_{(r_{x},k),(r_{x}+1,k)},
\end{equation}
where the local density reads
\begin{equation}
h^{(x)}_{(r_{x},k),(r_{x}+1,k)}=
\mathcal{P}_{(r_{x},k),(r_{x}+1,k)}^{(g)}\,
\frac{\partial}{\partial u_x}\mathbb{R}^{(2D,2S)}_{(r_{x},k),(r_{x}+1,k)}
   (\bm{u}_{0}; u_0; \bm{\zeta}; \bm{k}; \lambda, \vartheta).
\end{equation}
Analogously, for a fixed column $j$ we obtain the $y$‑direction Hamiltonian
\begin{equation}
H^{(y)}=\sum_{r_y=1}^{L_y}h^{(y)}_{(j,r_{y}),(j,r_{y}+1)},
\end{equation}
with local density
\begin{equation}\label{69}
h^{(y)}_{(j,r_{y}),(j,r_{y}+1)}=
\mathcal{P}_{(j,r_{y}),(j,r_{y}+1)}^{(g)}\,
\frac{\partial}{\partial u_y}\mathbb{R}^{(2D,2S)}_{(j,r_{y}),(j,r_{y}+1)}
   (\bm{u}_{0}; u_0; \bm{\zeta}; \bm{k}; \lambda, \vartheta).
\end{equation}

Denoting nearest‑neighbor pairs in the $x$ and $y$ directions by $\langle r,r^{\prime}\rangle_{x}$ and $\langle r,r^{\prime}\rangle_{y}$, respectively, and using the fermionic operators $c_{r\sigma},c_{r\sigma}^{\dagger}$ and $n_{r\sigma}=c_{r\sigma}^{\dagger}c_{r\sigma}$ ($\sigma=\uparrow,\downarrow$), the total Hamiltonian obtained from ~\eqref{66}--\eqref{69} takes the form of an anisotropic Hubbard model with additional orbital couplings:
\begin{equation}\label{70}
\begin{aligned}
H^{(2D)}&=H^{(x)}+H^{(y)}
        =\sum_{\langle r,r^{\prime}\rangle_{x}}h^{(x)}_{r,r^{\prime}}
         +\sum_{\langle r,r^{\prime}\rangle_{y}}h^{(y)}_{r,r^{\prime}}\\[2mm]
&=-t_x \sum_{\langle r, r+\hat{x}\rangle,\sigma}
        \bigl(c_{r\sigma}^\dagger c_{r+\hat{x},\sigma}+c_{r+\hat{x},\sigma}^\dagger c_{r\sigma}\bigr)
   -t_y \sum_{\langle r, r+\hat{y}\rangle,\sigma}
        \bigl(c_{r\sigma}^\dagger c_{r+\hat{y},\sigma}+c_{r+\hat{y},\sigma}^\dagger c_{r\sigma}\bigr)\\[1mm]
&\quad+U \sum_{r}\bigl(n_{r\uparrow}-\tfrac12\bigr)\bigl(n_{r\downarrow}-\tfrac12\bigr)
        +H_{\text{ext}}[\boldsymbol{\zeta},\boldsymbol{k}],\qquad \sigma=\uparrow,\downarrow .
\end{aligned}
\end{equation}
The hopping amplitudes and on‑site interaction are given by
\begin{equation}
\begin{aligned}
t_x &= -\langle \sigma,0|_{r,r+\hat{x}}\,
      \mathcal{P}_{r,r+\hat{x}}^{(g)}
      \Bigl.\frac{\partial}{\partial u_x}
      \mathbb{R}_{(r,r+\hat{x})}^{(2D,2s)}
      (\bm{u};u_0;\boldsymbol{\zeta};\bm{k};\lambda,\vartheta)\Bigr|_{\bm{u}=\bm{u}_0}
      |0,\sigma\rangle_{r,r+\hat{x}},\\[2mm]
t_y &= -\langle \sigma,0|_{r,r+\hat{y}}\,
      \mathcal{P}_{r,r+\hat{y}}^{(g)}
      \Bigl.\frac{\partial}{\partial u_y}
      \mathbb{R}_{(r,r+\hat{y})}^{(2D,2s)}
      (\bm{u};u_0;\boldsymbol{\zeta};\bm{k};\lambda,\vartheta)\Bigr|_{\bm{u}=\bm{u}_0}
      |0,\sigma\rangle_{r,r+\hat{y}},\\[2mm]
U   &= 2\bigl(\langle\uparrow\downarrow,0|-\langle\uparrow,0|-\langle\downarrow,0|+\langle0,0|\bigr)\,
      \mathcal{P}_{r,r+\hat{x}}^{(g)}
      \Bigl.\frac{\partial}{\partial u_x}
      \mathbb{R}_{(r,r+\hat{x})}^{(2D,2s)}
      (\bm{u};u_0;\boldsymbol{\zeta};\bm{k};\lambda,\vartheta)\Bigr|_{\bm{u}=\bm{u}_0}
      \begin{pmatrix}
      |\uparrow\downarrow,0\rangle\\|\uparrow,0\rangle\\|\downarrow,0\rangle\\|0,0\rangle
      \end{pmatrix}_{r,r+\hat{x}} .
\end{aligned}
\end{equation}
The term $H_{\text{ext}}[\boldsymbol{\zeta},\boldsymbol{k}]$ contains the orbital‑coupling contributions,
\begin{equation}\label{72}
\begin{aligned}
H_{\text{ext}}[\boldsymbol{\zeta},\bm{k}]
&=\sum_{\mu=x,y}\sum_{\langle r,r'\rangle_\mu}\sum_{\sigma}
   \Bigl(\delta t^{(\mu)}(\boldsymbol{\zeta},\bm{k})c_{r\sigma}^\dagger c_{r'\sigma}
        +\delta t^{(\mu)}(\boldsymbol{\zeta},\bm{k})^*c_{r'\sigma}^\dagger c_{r\sigma}\Bigr)\\
&\quad+\sum_{\mu=x,y}\sum_{\langle r,r'\rangle_\mu}\sum_{\sigma}
   \Bigl(\varepsilon_{r,\sigma}^{(\mu)}(\boldsymbol{\zeta},\bm{k})
        \bigl(n_{r\sigma}-\tfrac12\bigr)
        +\varepsilon_{r',\sigma}^{(\mu)}(\boldsymbol{\zeta},\bm{k})
        \bigl(n_{r'\sigma}-\tfrac12\bigr)\Bigr)\\
&\quad+\sum_{\mu=x,y}\sum_{\langle r,r'\rangle_\mu}
   \Bigl[\delta U_r^{(\mu)}(\boldsymbol{\zeta},\bm{k})
        \bigl(n_{r\uparrow}-\tfrac12\bigr)\bigl(n_{r\downarrow}-\tfrac12\bigr)
        +\delta U_{r'}^{(\mu)}(\boldsymbol{\zeta},\bm{k})
        \bigl(n_{r'\uparrow}-\tfrac12\bigr)\bigl(n_{r'\downarrow}-\tfrac12\bigr)\Bigr]\\
&\quad+\sum_{\mu=x,y}\sum_{\langle r,r'\rangle_\mu}
   V_{\text{nn}}^{(\mu)}(\boldsymbol{\zeta},\bm{k})
   \sum_{\sigma,\sigma'}\bigl(n_{r\sigma}-\tfrac12\bigr)\bigl(n_{r'\sigma'}-\tfrac12\bigr)\\
&\quad+\sum_{\mu=x,y}\sum_{\langle r,r'\rangle_\mu}C_0^{(\mu)}(\boldsymbol{\zeta},\bm{k}),
\end{aligned}
\end{equation}
whose explicit expressions are collected in Appendix~\ref{D}. Here $\hat{x}$ denotes the unit vector along the $x$‑direction. A key feature of the model is that the directional components do not commute; as shown in Appendix~\ref{F}, $[H^{(x)},H^{(y)}]\neq0$, confirming that the system is genuinely two‑dimensional and not a simple decoupled product of one‑dimensional chains.

The constructed Hamiltonian contains several well‑known limits. When the parameters are restricted so that $\partial_{u_y}\mathbb{R}^{(2D,2s)}=0$, all $y$‑direction couplings and extended coefficients with $\mu=y$ vanish. Further imposing $\delta t^{(x)}=\varepsilon^{(x)}=\delta U^{(x)}=V_{\text{nn}}^{(x)}=C_0^{(x)}=0$ eliminates the extended terms in the $x$‑direction, reducing the model to the standard one‑dimensional Hubbard chain:
\begin{equation}
H_{\text{1D Hubbard}} = -t \sum_{i,\sigma}
   \bigl(c_{i\sigma}^\dagger c_{i+1,\sigma}+c_{i+1,\sigma}^\dagger c_{i\sigma}\bigr)
   +U \sum_{i}\bigl(n_{i\uparrow}-\tfrac12\bigr)\bigl(n_{i\downarrow}-\tfrac12\bigr),
\end{equation}
where
\begin{equation}
\begin{aligned}
t &= -\langle \sigma,0|\,h_{i,i+1}^{(x)}\,|0,\sigma\rangle,\\[1mm]
U &= 2\bigl(\langle\uparrow\downarrow,0|-\langle\uparrow,0|-\langle\downarrow,0|+\langle0,0|\bigr)\,
      h_{i,i+1}^{(x)}
      \begin{pmatrix}
      |\uparrow\downarrow,0\rangle\\|\uparrow,0\rangle\\|\downarrow,0\rangle\\|0,0\rangle
      \end{pmatrix}_{i,i+1},\\[1mm]
h_{i,i+1}^{(x)} &=
\mathcal{P}_{i,i+1}^{(g)}\,
\Bigl.\frac{\partial}{\partial u_x}
\mathbb{R}_{i,i+1}^{(2D,2s)}
\bigl(\bm{u};\bm{u}_0;\boldsymbol{\zeta}^x,\boldsymbol{\zeta}^y;k_x,k_y;\lambda,\vartheta\bigr)
\Bigr|_{\substack{\bm{u}=\bm{u}_0\\
                   \partial_{u_y}\mathbb{R}^{(2D,2S)}=0,\,
                   \delta t^{(x)}=\varepsilon^{(x)}=\delta U^{(x)}=V_{\text{nn}}^{(x)}=C_0^{(x)}=0}}.
\end{aligned}
\end{equation}
Thus, the present construction encompasses both the integrable one‑dimensional Hubbard model and a genuinely coupled two‑dimensional extension with tunable anisotropic hopping, Hubbard interaction, and orbital couplings.

We now examine the scenario where $U \rightarrow 0$ and select the subspace that satisfies $\delta U^{(\mu)} = 0$ and $V_{\text{nn}}^{(\mu)} = 0$. Invoking \eqref{70}-\eqref{72} and \eqref{99}, we express the total Hamiltonian in this limit as the Free Fermion Hamiltonian:
\begin{equation}
H_{\text{free}}[\boldsymbol{\zeta}, \bm{k}] = \sum_{\sigma} \sum_{r,r'} A_{rr'}^{(\sigma)}(\boldsymbol{\zeta}, \bm{k}; 0) c_{r\sigma}^\dagger c_{r'\sigma} + E_0(\boldsymbol{\zeta}, \bm{k}),
\end{equation}
where\footnote{Here, the expression$\sum_{r' : \langle r,r'\rangle_\mu}$
signifies that we fix a specific lattice site $r$ and sum strictly over those sites $r'$ connected to $r$ via a nearest-neighbor bond $\langle r, r'\rangle_\mu$ along the $\mu$-direction. In other words, this summation enumerates all neighbors of $r$ in the $\mu$-direction. To illustrate this concretely, assuming the neighbor relation along the $x$-direction satisfies$\langle r, r'\rangle_x \Longleftrightarrow r' = r + \hat{x} \text{ or } r' = r - \hat{x},$
we find that for any internal lattice site $r$: $\sum_{r' : \langle r,r'\rangle_x} f(r, r') = f(r, r + \hat{x}) + f(r, r - \hat{x}).$
Should the site $r - \hat{x}$ exist outside the boundary, we naturally exclude the corresponding term from the summation.}
\begin{equation}
\begin{aligned}
    A_{rr'}^{(\sigma)}(\boldsymbol{\zeta}, \bm{k}; 0) &= \left(\sum_{\mu=x,y} \sum_{r' : \langle r, r' \rangle_\mu} \varepsilon_{r,\sigma}^{(\mu)}(\boldsymbol{\zeta}, \bm{k})\right)\delta_{rr'} \\&+ \sum_{\mu=x,y} \left[ \left(-t_\mu + \delta t^{(\mu)}(\boldsymbol{\zeta}, \bm{k})\right) \delta_{r', r+\hat{\mu}} + \left(-t_\mu + \delta t^{(\mu)}(\boldsymbol{\zeta}, \bm{k})\right)^* \delta_{r', r-\hat{\mu}} \right],
\end{aligned}
\end{equation}
\begin{equation}
E_0(\boldsymbol{\zeta}, \bm{k}) = -\frac{1}{2} \sum_{\mu=x,y} \sum_{\langle r,r'\rangle_{\mu,\sigma}} \left[ \varepsilon_{r,\sigma}^{(\mu)}(\boldsymbol{\zeta}, \bm{k}) + \varepsilon_{r',\sigma}^{(\mu)}(\boldsymbol{\zeta}, \bm{k}) \right] + \sum_{\mu=x,y} \sum_{\langle r,r'\rangle_\mu} C_0^{(\mu)}(\boldsymbol{\zeta}, \bm{k}).
\end{equation}

To summarize, we have found: (i) a construction of a strictly solvable two-dimensional lattice model defined by a global R-operator $\mathbb{R}_{ab}^{(2D,2S)}$ \eqref{13} that satisfies the Yang-Baxter equation\eqref{14}, achieved by anisotropizing free-fermion L-operators and incorporating a spin-charge coupling factor, and (ii) the exact solution of the model using the nested algebraic Bethe ansatz, which yields the eigenvalues of the transfer matrix along with the \eqref{55} first- and \eqref{58} second-level Bethe equations, as well as the explicit derivation of the resulting Hamiltonian \eqref{70} describing a layered Hubbard model with anisotropic staggered couplings.
\section*{Declaration of competing interest}
The authors declared that they have no conflicts of interest to this work. 

\section*{Acknowledgment}
This work is supported by the developing Project of Science and Technology of Jilin Province (20240402042GH). 

\section*{Data availability}
Data will be made available on request.

\appendix
\renewcommand\thefigure{\Alph{section}\arabic{figure}} 
\renewcommand\theequation{\Alph{section}\arabic{equation}} 

\section{Outer even bilinear basis on the two-site fermionic Hilbert space}\label{A}
\setcounter{figure}{0}
\setcounter{equation}{0}
We verify the commutation relations of the spin-charge coupling operator with the basis elements:
\begin{equation}\label{78}
\begin{aligned}
\mathcal{O}_a\mathcal{O}_b n_{a\sigma} (\mathcal{O}_a\mathcal{O}_b)^{-1} &= n_{a\sigma}, \quad \mathcal{O}_a\mathcal{O}_b n_{b\sigma} (\mathcal{O}_a\mathcal{O}_b)^{-1} = n_{b\sigma}, \\
\mathcal{O}_a\mathcal{O}_b \Delta_{ab}^{(\sigma)} (\mathcal{O}_a\mathcal{O}_b)^{-1} &= \mathcal{O}_a\mathcal{O}_b (c_{a\sigma}^{\dagger}c_{b\sigma}) (\mathcal{O}_a\mathcal{O}_b)^{-1} = (-c_{a\sigma}^{\dagger})(-c_{b\sigma}) = \Delta_{ab}^{(\sigma)}, \\
\mathcal{O}_a\mathcal{O}_b \Delta_{ba}^{(\sigma)} (\mathcal{O}_a\mathcal{O}_b)^{-1} &= \Delta_{ba}^{(\sigma)}, \\
\mathcal{O}_a\mathcal{O}_b \widetilde{\Delta}_{ab}^{(+,\sigma)} (\mathcal{O}_a\mathcal{O}_b)^{-1} &= \mathcal{O}_a\mathcal{O}_b (c_{a\sigma}^{\dagger}c_{b\sigma}^{\dagger}) (\mathcal{O}_a\mathcal{O}_b)^{-1} = (-c_{a\sigma}^{\dagger})(-c_{b\sigma}^{\dagger}) = \widetilde{\Delta}_{ab}^{(+,\sigma)}, \\
\mathcal{O}_a\mathcal{O}_b \widetilde{\Delta}_{ab}^{(\sigma)} (\mathcal{O}_a\mathcal{O}_b)^{-1} &= \widetilde{\Delta}_{ab}^{(\sigma)}.
\end{aligned}
\end{equation}
\section{Jacobi-elliptic expansion of the global R-matrix components}\label{B}
\setcounter{figure}{0}
\setcounter{equation}{0}
We denote the spectral parameters as two-dimensional vectors $\boldsymbol{u},\boldsymbol{v}$, the external field as the vector $\boldsymbol{\zeta}$, and the elliptic modulus as the vector $\boldsymbol{k}$, namely:
\begin{equation}
    \boldsymbol{u} = (u_x, u_y),\quad \boldsymbol{v} = (v_x, v_y),\quad \boldsymbol{\zeta} = (\zeta_x, \zeta_y),\quad \boldsymbol{k} = (k_x, k_y).
\end{equation}
We define the unit vector $\hat{\boldsymbol{d}} = \frac{1}{\sqrt{2}}(1,1),$ and thereby define the scalar argument of the spectral difference as:
\begin{equation}
u_{ab}(\bm{u},\bm{v}) := \hat{\bm{d}} \cdot (\bm{u} - \bm{v}) = \frac{1}{\sqrt{2}} \left[ (u_1 - v_1) + (u_2 - v_2) \right].
\end{equation}
We define the external field scalar $\zeta$ as:
\begin{equation}
\zeta(\bm{\zeta}) := \hat{\bm{d}} \cdot \bm{\zeta} = \frac{1}{\sqrt{2}} (\zeta_1 + \zeta_2).
\end{equation}
We reduce the elliptic modulus vector to a scalar modulus $\kappa$:
\begin{equation}
\kappa := \|\bm{k}\| = \sqrt{k_1^2 + k_2^2},
\end{equation}
and subsequently define the Jacobi combinations:
\begin{equation}
\begin{aligned}
E &= e(u_{ab}; \boldsymbol{k}) = \operatorname{cn}(u_{ab}; \boldsymbol{k}) + i \operatorname{sn}(u_{ab}; \boldsymbol{k}), \\
Z &= e(\zeta; \boldsymbol{k}) = \operatorname{cn}(\zeta; \boldsymbol{k}) + i \operatorname{sn}(\zeta; \boldsymbol{k}), \\
S &= \operatorname{sn}\bigl(\frac{u_{ab}}{2}; \boldsymbol{k}\bigr), \quad s = \operatorname{sn}(\zeta; \boldsymbol{k}).
\end{aligned}
\end{equation}
Here, we denote the normalization factor as $\rho = \rho(u_{ab},\zeta)$ and the effective modulus extracted from the vector $\boldsymbol{k}$ as $\kappa$. All coefficients depend solely on these scalar combinations.

We define four ``diagonal'' coefficients:
\begin{equation}
\begin{aligned}
A &:= \rho(E - Z^2), \\
B &:= \rho(Z - EZ) = \rho Z(1 - E), \\
C &:= \rho(E Z^2 - 1),
\end{aligned}
\end{equation}
We further define two ``jump'' and two ``pairing'' amplitudes:
\begin{equation}
\begin{aligned}
P &:= \rho(1 - E) S^{-1} Z s, \\
U &:= i \kappa \rho S(1 + E) Z^2 s^2, \\
V &:= -i \kappa \rho S(1 + E).
\end{aligned}
\end{equation}
We then observe that $\mathbb{R}^{\alpha\beta}_{\quad\gamma\delta}$ satisfies the following structure:
\begin{equation}
\begin{aligned}
\mathbb{R}_{\quad11}^{11} &= A \cdot A = \rho^2 \left( E - Z^2 \right)^2, & \mathbb{R}_{\quad22}^{11} &= V \cdot A = -i \kappa \rho^2 S \left( 1 + E \right) \left( E - Z^2 \right), \\
\mathbb{R}_{\quad33}^{11} &= A \cdot V = -i \kappa \rho^2 S \left( 1 + E \right) \left( E - Z^2 \right), & \mathbb{R}_{\quad44}^{11} &= V \cdot V = -\kappa^2 \rho^2 S^2 \left( 1 + E \right)^2, \\
\mathbb{R}_{\quad12}^{12} &= B \cdot A = \rho^2 \left( Z - EZ \right) \left( E - Z^2 \right), & \mathbb{R}_{\quad34}^{12} &= B \cdot V = -i \kappa \rho^2 S \left( 1 + E \right) \left( Z - EZ \right), \\
\mathbb{R}_{\quad21}^{12} &= P \cdot A = \rho^2 \left( 1 - E \right) S^{-1} Z s \left( E - Z^2 \right), & \mathbb{R}_{\quad43}^{12} &= P \cdot V = -i \kappa \rho^2 \left( 1 - E \right) Z s \left( 1 + E \right), \\
\mathbb{R}_{\quad13}^{13} &= A \cdot B = \rho^2 \left( E - Z^2 \right) \left( Z - EZ \right), & \mathbb{R}_{\quad31}^{13} &= A \cdot P = \rho^2 \left( E - Z^2 \right) \left( 1 - E \right) S^{-1} Z s, \\
\mathbb{R}_{\quad24}^{13} &= V \cdot B = -i \kappa \rho^2 S \left( 1 + E \right) \left( Z - EZ \right), & \mathbb{R}_{\quad42}^{13} &= V \cdot P = -i \kappa \rho^2 \left( 1 - E \right) Z s \left( 1 + E \right), \\
\mathbb{R}_{\quad14}^{14} &= B \cdot B = \rho^2 \left( Z - EZ \right)^2, & \mathbb{R}_{\quad32}^{14} &= B \cdot P = \rho^2 \left( Z - EZ \right) \left( 1 - E \right) S^{-1} Z s, \\
\mathbb{R}_{\quad23}^{14} &= P \cdot B = \rho^2 \left( 1 - E \right) S^{-1} Z s \left( Z - EZ \right), & \mathbb{R}_{\quad41}^{14} &= P \cdot P = \rho^2 \left( 1 - E \right)^2 S^{-2} Z^2 s^2, \\
\mathbb{R}_{\quad21}^{21} &= B \cdot A = \rho^2 \left( Z - EZ \right) \left( E - Z^2 \right), & \mathbb{R}_{\quad43}^{21} &= B \cdot V = -i \kappa \rho^2 S \left( 1 + E \right) \left( Z - EZ \right), \\
\mathbb{R}_{\quad12}^{21} &= P \cdot A = \rho^2 \left( 1 - E \right) S^{-1} Z s \left( E - Z^2 \right), & \mathbb{R}_{\quad34}^{21} &= P \cdot V = -i \kappa \rho^2 \left( 1 - E \right) Z s \left( 1 + E \right), \\
\mathbb{R}_{\quad22}^{22} &= C \cdot A = \rho^2 \left( EZ^2 - 1 \right) \left( E - Z^2 \right), & \mathbb{R}_{\quad44}^{22} &= C \cdot V = -i \kappa \rho^2 S \left( 1 + E \right) \left( EZ^2 - 1 \right), \\
\mathbb{R}_{\quad11}^{22} &= U \cdot A = i \kappa \rho^2 S \left( 1 + E \right) Z^2 s^2 \left( E - Z^2 \right), & \mathbb{R}_{\quad33}^{22} &= U \cdot V = -\kappa^2 \rho^2 S^2 \left( 1 + E \right)^2 Z^2 s^2, \\
\mathbb{R}_{\quad23}^{23} &= B \cdot B = \rho^2 \left( Z - EZ \right)^2, & \mathbb{R}_{\quad41}^{23} &= B \cdot P = \rho^2 \left( Z - EZ \right) \left( 1 - E \right) S^{-1} Z s, \\
\mathbb{R}_{\quad14}^{23} &= P \cdot B = \rho^2 \left( 1 - E \right) S^{-1} Z s \left( Z - EZ \right), & \mathbb{R}_{\quad32}^{23} &= P \cdot P = \rho^2 \left( 1 - E \right)^2 S^{-2} Z^2 s^2, \\
\mathbb{R}_{\quad24}^{24} &= C \cdot B = \rho^2 \left( EZ^2 - 1 \right) \left( Z - EZ \right), & \mathbb{R}_{\quad42}^{24} &= C \cdot P = \rho^2 \left( EZ^2 - 1 \right) \left( 1 - E \right) S^{-1} Z s, \\
\mathbb{R}_{\quad13}^{24} &= U \cdot B = i \kappa \rho^2 S \left( 1 + E \right) Z^2 s^2 \left( Z - EZ \right), & \mathbb{R}_{\quad31}^{24} &= U \cdot P = i \kappa \rho^2 \left( 1 + E \right) \left( 1 - E \right) Z^2 s^2, \\
\mathbb{R}_{\quad31}^{31} &= A \cdot B = \rho^2 \left( E - Z^2 \right) \left( Z - EZ \right), & \mathbb{R}_{\quad13}^{31} &= A \cdot P = \rho^2 \left( E - Z^2 \right) \left( 1 - E \right) S^{-1} Z s, \\
\mathbb{R}_{\quad42}^{31} &= V \cdot B = -i \kappa \rho^2 S \left( 1 + E \right) \left( Z - EZ \right), & \mathbb{R}_{\quad24}^{31} &= V \cdot P = -i \kappa \rho^2 \left( 1 + E \right) \left( 1 - E \right) Z s, \\
\mathbb{R}_{\quad32}^{32} &= B \cdot B = \rho^2 \left( Z - EZ \right)^2, & \mathbb{R}_{\quad14}^{32} &= B \cdot P = \rho^2 \left( Z - EZ \right) \left( 1 - E \right) S^{-1} Z s, \\
\mathbb{R}_{\quad41}^{32} &= P \cdot B = \rho^2 \left( 1 - E \right) S^{-1} Z s \left( Z - EZ \right), & \mathbb{R}_{\quad23}^{32} &= P \cdot P = \rho^2 \left( 1 - E \right)^2 S^{-2} Z^2 s^2, \\
\mathbb{R}_{\quad33}^{33} &= A \cdot C = \rho^2 \left( E - Z^2 \right) \left( EZ^2 - 1 \right), & \mathbb{R}_{\quad11}^{33} &= A \cdot U = i \kappa \rho^2 S \left( 1 + E \right) Z^2 s^2 \left( E - Z^2 \right). \\
\end{aligned}
\end{equation}
\begin{equation}
    \begin{aligned}
        \mathbb{R}_{\quad44}^{33} &= V \cdot C = -i \kappa \rho^2 S \left( 1 + E \right) \left( EZ^2 - 1 \right), & \mathbb{R}_{\quad22}^{33} &= V \cdot U = -\kappa^2 \rho^2 S^2 \left( 1 + E \right)^2 Z^2 s^2, \\
\mathbb{R}_{\quad34}^{34} &= B \cdot C = \rho^2 \left( Z - EZ \right) \left( EZ^2 - 1 \right), & \mathbb{R}_{\quad12}^{34} &= B \cdot U = i \kappa \rho^2 S \left( 1 + E \right) Z^2 s^2 \left( Z - EZ \right), \\
\mathbb{R}_{\quad43}^{34} &= P \cdot C = \rho^2 \left( 1 - E \right) S^{-1} Z s \left( EZ^2 - 1 \right), & \mathbb{R}_{\quad21}^{34} &= P \cdot U = i \kappa \rho^2 \left( 1 - E \right) Z^2 s^2 \left( 1 + E \right), \\
\mathbb{R}_{\quad41}^{41} &= B \cdot B = \rho^2 \left( Z - EZ \right)^2, & \mathbb{R}_{\quad23}^{41} &= B \cdot P = \rho^2 \left( Z - EZ \right) \left( 1 - E \right) S^{-1} Z s, \\
\mathbb{R}_{\quad32}^{41} &= P \cdot B = \rho^2 \left( 1 - E \right) S^{-1} Z s \left( Z - EZ \right), & \mathbb{R}_{\quad14}^{41} &= P \cdot P = \rho^2 \left( 1 - E \right)^2 S^{-2} Z^2 s^2, \\
\mathbb{R}_{\quad42}^{42} &= C \cdot B = \rho^2 \left( EZ^2 - 1 \right) \left( Z - EZ \right), & \mathbb{R}_{\quad24}^{42} &= C \cdot P = \rho^2 \left( EZ^2 - 1 \right) \left( 1 - E \right) S^{-1} Z s, \\
\mathbb{R}_{\quad31}^{42} &= U \cdot B = i \kappa \rho^2 S \left( 1 + E \right) Z^2 s^2 \left( Z - EZ \right), & \mathbb{R}_{\quad13}^{42} &= U \cdot P = i \kappa \rho^2 \left( 1 + E \right) \left( 1 - E \right) Z^2 s^2, \\
\mathbb{R}_{\quad43}^{43} &= B \cdot C = \rho^2 \left( Z - EZ \right) \left( EZ^2 - 1 \right), & \mathbb{R}_{\quad21}^{43} &= B \cdot U = i \kappa \rho^2 S \left( 1 + E \right) Z^2 s^2 \left( Z - EZ \right), \\
\mathbb{R}_{\quad34}^{43} &= P \cdot C = \rho^2 \left( 1 - E \right) S^{-1} Z s \left( EZ^2 - 1 \right), & \mathbb{R}_{\quad13}^{43} &= P \cdot U = i \kappa \rho^2 \left( 1 - E \right) Z^2 s^2 \left( 1 + E \right), \\
\mathbb{R}_{\quad44}^{44} &= C \cdot C = \rho^2 \left( EZ^2 - 1 \right)^2, & \mathbb{R}_{\quad22}^{44} &= C \cdot U = i \kappa \rho^2 S \left( 1 + E \right) Z^2 s^2 \left( EZ^2 - 1 \right), \\
\mathbb{R}_{\quad33}^{44} &= U \cdot C = i \kappa \rho^2 S \left( 1 + E \right) Z^2 s^2 \left( EZ^2 - 1 \right), & \mathbb{R}_{\quad11}^{44} &= U \cdot U = -\kappa^2 \rho^2 S^2 \left( 1 + E \right)^2 Z^4 s^4.
    \end{aligned}
\end{equation}
\section{Component expansion of the layer-to-layer monodromy matrix elements}\label{C}
\setcounter{figure}{0}
\setcounter{equation}{0}
Based on  \eqref{15}-\eqref{18}, we readily derive each component of $T_{\alpha\beta}(\bm{u})$ in  \eqref{25}:
\begin{equation}
T_{11}(\bm{u}) = \sum_{\alpha_1,\dots,\alpha_{L_x-1}=1}^{4} \sum_{\beta_{1,k},\dots,\beta_{L_x,k}=1}^{4} \sum_{\delta_{1,k},\dots,\delta_{L_x,k}=1}^{4} \left[ \prod_{j=1}^{L_x} \mathbb{R}_{\quad\alpha_{j-1} \delta_{j,k}}^{\alpha_j \beta_{j,k}}(\bm{u}, \bm{p}_{j,k}; \bm{\zeta}_{j,k}; \bm{k}) \right] \bigotimes_{j=1}^{L_x} e_{\beta_{j,k} \delta_{j,k}}^{(j,k)} \bigg|_{\alpha_{L_x}=1,\,\alpha_0=1},
\end{equation}
\begin{equation}
T_{12}(\bm{u}) = \sum_{\alpha_1,\dots,\alpha_{L_x-1}=1}^{4} \sum_{\beta_{1,k},\dots,\beta_{L_x,k}=1}^{4} \sum_{\delta_{1,k},\dots,\delta_{L_x,k}=1}^{4} \left[ \prod_{j=1}^{L_x} \mathbb{R}_{\quad\alpha_{j-1} \delta_{j,k}}^{\alpha_j \beta_{j,k}}(\bm{u}, \bm{p}_{j,k}; \bm{\zeta}_{j,k}; \bm{k}) \right] \bigotimes_{j=1}^{L_x} e_{\beta_{j,k} \delta_{j,k}}^{(j,k)} \bigg|_{\alpha_{L_x}=1,\,\alpha_0=2},
\end{equation}
\begin{equation}
T_{13}(\bm{u}) = \sum_{\alpha_1,\dots,\alpha_{L_x-1}=1}^{4} \sum_{\beta_{1,k},\dots,\beta_{L_x,k}=1}^{4} \sum_{\delta_{1,k},\dots,\delta_{L_x,k}=1}^{4} \left[ \prod_{j=1}^{L_x} \mathbb{R}_{\quad\alpha_{j-1} \delta_{j,k}}^{\alpha_j \beta_{j,k}}(\bm{u}, \bm{p}_{j,k}; \bm{\zeta}_{j,k}; \bm{k}) \right] \bigotimes_{j=1}^{L_x} e_{\beta_{j,k} \delta_{j,k}}^{(j,k)}  \bigg|_{\alpha_{L_x}=1,\,\alpha_0=3},
\end{equation}
\begin{equation}
T_{14}(\bm{u}) = \sum_{\alpha_1,\dots,\alpha_{L_x-1}=1}^{4} \sum_{\beta_{1,k},\dots,\beta_{L_x,k}=1}^{4} \sum_{\delta_{1,k},\dots,\delta_{L_x,k}=1}^{4} \left[ \prod_{j=1}^{L_x} \mathbb{R}_{\quad\alpha_{j-1} \delta_{j,k}}^{\alpha_j \beta_{j,k}}(\bm{u}, \bm{p}_{j,k}; \bm{\zeta}_{j,k}; \bm{k}) \right] \bigotimes_{j=1}^{L_x} e_{\beta_{j,k} \delta_{j,k}}^{(j,k)} \bigg|_{\alpha_{L_x}=1,\,\alpha_0=4},
\end{equation}
\begin{equation}
T_{21}(\bm{u}) = \sum_{\alpha_1,\dots,\alpha_{L_x-1}=1}^{4} \sum_{\beta_{1,k},\dots,\beta_{L_x,k}=1}^{4} \sum_{\delta_{1,k},\dots,\delta_{L_x,k}=1}^{4} \left[ \prod_{j=1}^{L_x} \mathbb{R}_{\quad\alpha_{j-1} \delta_{j,k}}^{\alpha_j \beta_{j,k}}(\bm{u}, \bm{p}_{j,k}; \bm{\zeta}_{j,k}; \bm{k}) \right] \bigotimes_{j=1}^{L_x} e_{\beta_{j,k} \delta_{j,k}}^{(j,k)}  \bigg|_{\alpha_{L_x}=2,\,\alpha_0=1},
\end{equation}
\begin{equation}
T_{22}(\bm{u}) = \sum_{\alpha_1,\dots,\alpha_{L_x-1}=1}^{4} \sum_{\beta_{1,k},\dots,\beta_{L_x,k}=1}^{4} \sum_{\delta_{1,k},\dots,\delta_{L_x,k}=1}^{4} \left[ \prod_{j=1}^{L_x} \mathbb{R}_{\quad\alpha_{j-1} \delta_{j,k}}^{\alpha_j \beta_{j,k}}(\bm{u}, \bm{p}_{j,k}; \bm{\zeta}_{j,k}; \bm{k}) \right] \bigotimes_{j=1}^{L_x} e_{\beta_{j,k} \delta_{j,k}}^{(j,k)} \bigg|_{\alpha_{L_x}=2,\,\alpha_0=2},
\end{equation}
\begin{equation}
T_{23}(\bm{u}) = \sum_{\alpha_1,\dots,\alpha_{L_x-1}=1}^{4} \sum_{\beta_{1,k},\dots,\beta_{L_x,k}=1}^{4} \sum_{\delta_{1,k},\dots,\delta_{L_x,k}=1}^{4} \left[ \prod_{j=1}^{L_x} \mathbb{R}_{\quad\alpha_{j-1} \delta_{j,k}}^{\alpha_j \beta_{j,k}}(\bm{u}, \bm{p}_{j,k}; \bm{\zeta}_{j,k}; \bm{k}) \right] \bigotimes_{j=1}^{L_x} e_{\beta_{j,k} \delta_{j,k}}^{(j,k)} \bigg|_{\alpha_{L_x}=2,\,\alpha_0=3},
\end{equation}
\begin{equation}
T_{31}(\bm{u}) = \sum_{\alpha_1,\dots,\alpha_{L_x-1}=1}^{4} \sum_{\beta_{1,k},\dots,\beta_{L_x,k}=1}^{4} \sum_{\delta_{1,k},\dots,\delta_{L_x,k}=1}^{4} \left[ \prod_{j=1}^{L_x} \mathbb{R}_{\quad\alpha_{j-1} \delta_{j,k}}^{\alpha_j \beta_{j,k}}(\bm{u}, \bm{p}_{j,k}; \bm{\zeta}_{j,k}; \bm{k}) \right] \bigotimes_{j=1}^{L_x} e_{\beta_{j,k} \delta_{j,k}}^{(j,k)} \bigg|_{\alpha_{L_x}=3,\,\alpha_0=1},
\end{equation}
\begin{equation}
T_{32}(\bm{u}) = \sum_{\alpha_1,\dots,\alpha_{L_x-1}=1}^{4} \sum_{\beta_{1,k},\dots,\beta_{L_x,k}=1}^{4} \sum_{\delta_{1,k},\dots,\delta_{L_x,k}=1}^{4} \left[ \prod_{j=1}^{L_x} \mathbb{R}_{\quad\alpha_{j-1} \delta_{j,k}}^{\alpha_j \beta_{j,k}}(\bm{u}, \bm{p}_{j,k}; \bm{\zeta}_{j,k}; \bm{k}) \right] \bigotimes_{j=1}^{L_x} e_{\beta_{j,k} \delta_{j,k}}^{(j,k)} \bigg|_{\alpha_{L_x}=3,\,\alpha_0=2},
\end{equation}
\begin{equation}
T_{33}(\bm{u}) = \sum_{\alpha_1,\dots,\alpha_{L_x-1}=1}^{4} \sum_{\beta_{1,k},\dots,\beta_{L_x,k}=1}^{4} \sum_{\delta_{1,k},\dots,\delta_{L_x,k}=1}^{4} \left[ \prod_{j=1}^{L_x} \mathbb{R}_{\quad\alpha_{j-1} \delta_{j,k}}^{\alpha_j \beta_{j,k}}(\bm{u}, \bm{p}_{j,k}; \bm{\zeta}_{j,k}; \bm{k}) \right] \bigotimes_{j=1}^{L_x} e_{\beta_{j,k} \delta_{j,k}}^{(j,k)}  \bigg|_{\alpha_{L_x}=3,\,\alpha_0=3},
\end{equation}
\begin{equation}
T_{34}(\bm{u}) = \sum_{\alpha_1,\dots,\alpha_{L_x-1}=1}^{4} \sum_{\beta_{1,k},\dots,\beta_{L_x,k}=1}^{4} \sum_{\delta_{1,k},\dots,\delta_{L_x,k}=1}^{4} \left[ \prod_{j=1}^{L_x} \mathbb{R}_{\quad\alpha_{j-1} \delta_{j,k}}^{\alpha_j \beta_{j,k}}(\bm{u}, \bm{p}_{j,k}; \bm{\zeta}_{j,k}; \bm{k}) \right] \bigotimes_{j=1}^{L_x} e_{\beta_{j,k} \delta_{j,k}}^{(j,k)} \bigg|_{\alpha_{L_x}=3,\,\alpha_0=4},
\end{equation}
\begin{equation}
T_{41}(\bm{u}) = \sum_{\alpha_1,\dots,\alpha_{L_x-1}=1}^{4} \sum_{\beta_{1,k},\dots,\beta_{L_x,k}=1}^{4} \sum_{\delta_{1,k},\dots,\delta_{L_x,k}=1}^{4} \left[ \prod_{j=1}^{L_x} \mathbb{R}_{\quad\alpha_{j-1} \delta_{j,k}}^{\alpha_j \beta_{j,k}}(\bm{u}, \bm{p}_{j,k}; \bm{\zeta}_{j,k}; \bm{k}) \right] \bigotimes_{j=1}^{L_x} e_{\beta_{j,k} \delta_{j,k}}^{(j,k)} \bigg|_{\alpha_{L_x}=4,\,\alpha_0=1},
\end{equation}
\begin{equation}
T_{42}(\bm{u}) = \sum_{\alpha_1,\dots,\alpha_{L_x-1}=1}^{4} \sum_{\beta_{1,k},\dots,\beta_{L_x,k}=1}^{4} \sum_{\delta_{1,k},\dots,\delta_{L_x,k}=1}^{4} \left[ \prod_{j=1}^{L_x} \mathbb{R}_{\quad\alpha_{j-1} \delta_{j,k}}^{\alpha_j \beta_{j,k}}(\bm{u}, \bm{p}_{j,k}; \bm{\zeta}_{j,k}; \bm{k}) \right] \bigotimes_{j=1}^{L_x} e_{\beta_{j,k} \delta_{j,k}}^{(j,k)} \bigg|_{\alpha_{L_x}=4,\,\alpha_0=2},
\end{equation}
\begin{equation}
T_{43}(\bm{u}) = \sum_{\alpha_1,\dots,\alpha_{L_x-1}=1}^{4} \sum_{\beta_{1,k},\dots,\beta_{L_x,k}=1}^{4} \sum_{\delta_{1,k},\dots,\delta_{L_x,k}=1}^{4} \left[ \prod_{j=1}^{L_x} \mathbb{R}_{\quad\alpha_{j-1} \delta_{j,k}}^{\alpha_j \beta_{j,k}}(\bm{u}, \bm{p}_{j,k}; \bm{\zeta}_{j,k}; \bm{k}) \right] \bigotimes_{j=1}^{L_x} e_{\beta_{j,k} \delta_{j,k}}^{(j,k)} \bigg|_{\alpha_{L_x}=4,\,\alpha_0=3},
\end{equation}
\begin{equation}
T_{44}(\bm{u}) = \sum_{\alpha_1,\dots,\alpha_{L_x-1}=1}^{4} \sum_{\beta_{1,k},\dots,\beta_{L_x,k}=1}^{4} \sum_{\delta_{1,k},\dots,\delta_{L_x,k}=1}^{4} \left[ \prod_{j=1}^{L_x} \mathbb{R}_{\quad\alpha_{j-1} \delta_{j,k}}^{\alpha_j \beta_{j,k}}(\bm{u}, \bm{p}_{j,k}; \bm{\zeta}_{j,k}; \bm{k}) \right] \bigotimes_{j=1}^{L_x} e_{\beta_{j,k} \delta_{j,k}}^{(j,k)} \bigg|_{\alpha_{L_x}=4,\,\alpha_0=4}.
\end{equation}
\section{Explicit expressions for the extended Hamiltonian parameters}\label{D}
\setcounter{figure}{0}
\setcounter{equation}{0}
In \eqref{72}, we have:
\begin{equation}\label{99}
    \begin{aligned}
        &C_0^{(\mu)}(\boldsymbol{\zeta}, \bm{k}) = \left. \frac{1}{16} \text{Tr}\left[ \mathcal{P}_{r',r}^{(g)} \partial_{u_\mu} \mathbb{R}_{r',r}^{(2D,2s)} \right] \right|_{\bm{u}=\bm{u}_0},\\
        &\varepsilon_{r,\sigma}^{(\mu)}(\boldsymbol{\zeta}, \bm{k}) =\left. \frac{1}{4} \text{Tr}\left[ \left(n_{r\sigma} - \frac{1}{2}\right) \mathcal{P}_{r',r}^{(g)} \partial_{u_\mu} \mathbb{R}_{r',r}^{(2D,2s)} \right] \right|_{\bm{u}=\bm{u}_0},\\
        &\varepsilon_{r',\sigma}^{(\mu)}(\boldsymbol{\zeta}, \bm{k}) = \left. \frac{1}{4} \text{Tr}\left[ \left(n_{r'\sigma} - \frac{1}{2}\right) \mathcal{P}_{r',r}^{(g)} \partial_{u_\mu} \mathbb{R}_{r',r}^{(2D,2s)} \right] \right|_{\bm{u}=\bm{u}_0},\\
        &\delta U_r^{(\mu)}(\boldsymbol{\zeta}, \bm{k})=\left. \text{Tr}\left[ \left(n_{r\uparrow} - \frac{1}{2}\right)\left(n_{r\downarrow} - \frac{1}{2}\right) \mathcal{P}_{r',r}^{(g)} \partial_{u_\mu} \mathbb{R}_{r',r}^{(2D,2s)} \right] \right|_{\bm{u}=\bm{u}_0} - \frac{U}{2},\\
        &\delta U_{r'}^{(\mu)}(\boldsymbol{\zeta}, \bm{k}) = \left. \text{Tr}\left[ \left(n_{r'\uparrow} - \frac{1}{2}\right)\left(n_{r'\downarrow} - \frac{1}{2}\right) \mathcal{P}_{r',r}^{(g)} \partial_{u_\mu} \mathbb{R}_{r',r}^{(2D,2s)} \right] \right|_{\bm{u}=\bm{u}_0} - \frac{U}{2},\\
        &V_{\text{nn}}^{(\mu)}(\boldsymbol{\zeta}, \bm{k})=\left. \frac{1}{4} \text{Tr}\left[ \sum_{\sigma,\sigma'} \left(n_{r\sigma} - \frac{1}{2}\right)\left(n_{r'\sigma'} - \frac{1}{2}\right) \mathcal{P}_{r',r}^{(g)} \partial_{u_\mu} \mathbb{R}_{r',r}^{(2D,2s)} \right] \right|_{\bm{u}=\bm{u}_0},\\
        &\delta t^{(\mu)}(\boldsymbol{\zeta}, \bm{k})=\left. -\frac{1}{16} \text{Tr}\left[ \sum_{\sigma} \left(c_{r\sigma}^\dagger c_{r'\sigma} + c_{r'\sigma}^\dagger c_{r\sigma}\right) \mathcal{P}_{r',r}^{(g)} \partial_{u_\mu} \mathbb{R}_{r',r}^{(2D,2s)} \right] \right|_{\bm{u}=\bm{u}_0} - t_\mu.
    \end{aligned}
\end{equation}
where $\mu=x,y$, and $\sigma = \uparrow, \downarrow$.

\section{Exact diagonalization validation on finite lattices}\label{E}
\setcounter{figure}{0}
\setcounter{equation}{0}

\begin{figure*}[t]
	\centering
	\includegraphics[width=1.05\textwidth]{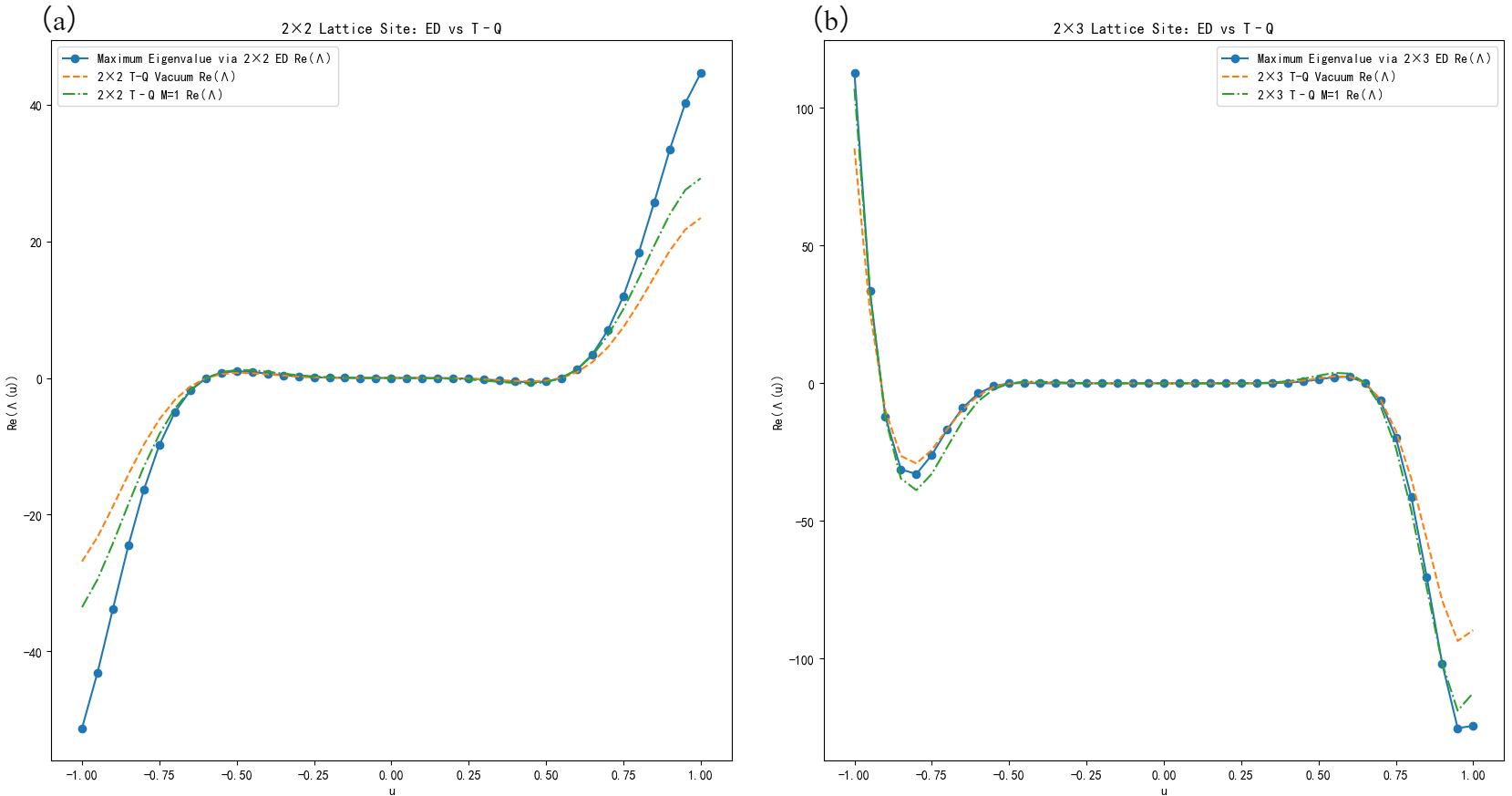}
	\caption{Benchmarking Bethe–TQ eigenvalues against exact diagonalization for small lattices governed by the constructed Hamiltonian. \textbf{(a)} $2 \times 2$ lattice: real part of the largest transfer-matrix eigenvalue $\Lambda(u)$ obtained from $2 \times 2$ ED (dots) and from the vacuum and one-particle Bethe-TQ solutions (dashed and dash-dotted lines).\textbf{(b)} $2 \times 3$ lattice: real part of the largest transfer-matrix eigenvalue $\Lambda(u)$ obtained from $2 \times 3$ ED (dots) and from the vacuum and one-particle Bethe-TQ solutions (dashed and dash-dotted lines).
   }
	\label{F1}
\end{figure*}

To validate the integrable construction developed in the previous sections, we compare the largest eigenvalue of the transfer matrix obtained from exact diagonalization (ED) on finite lattices with the analytic predictions of the nested Bethe ansatz. The comparison is carried out for two small lattices: a $2\times2$ lattice ($L_x=L_y=2$, total sites $N=4$) and a $2\times3$ lattice ($L_x=2,L_y=3$, $N=6$). The results are displayed in Fig.~\ref{F1}, where we plot the real part of the eigenvalue $\Lambda(u)$ as a function of a one‑dimensional cut through the spectral parameter space.

The transfer matrix $\tau^{(2D,2S)}(\boldsymbol{u})$ is built following the steps outlined in Sections~\ref{b}--\ref{c}. Starting from the anisotropic spectral vectors and deformed free‑fermion $L$‑operators introduced in ~\eqref{7}--\eqref{9}, the global two‑spin R‑operator $\mathbb{R}_{ab}^{(2D,2S)}$ is defined in ~\eqref{13}. Restricting one leg of this operator to a physical site yields the local $L$‑operators $L_{a,(j,k)}$ (~\eqref{15}--\eqref{17}), which are then assembled into horizontal and vertical monodromies $T_{a|k}^{(x)}(\boldsymbol{u})$ and $T_{a|j}^{(y)}(\boldsymbol{u})$. The full layer‑to‑layer monodromy $\Gamma_a(\boldsymbol{u})$ is formed by multiplying these directional monodromies (~\eqref{18}), and the transfer matrix is obtained as the graded trace $\tau^{(2D,2S)}(\boldsymbol{u}) = \operatorname{Str}_a \Gamma_a(\boldsymbol{u})$ (~\eqref{19}). For the numerical evaluation we choose homogeneous lattice parameters $\boldsymbol{p}_{jk}=\boldsymbol{0}$ and $\boldsymbol{\zeta}_{jk}=\boldsymbol{\zeta}$, fix the modulus $\boldsymbol{k}$, and compute for each real $u$ the dominant eigenvalue $\Lambda_{\text{ED}}(u)$ by ED of the $4^{L_xL_y}$‑dimensional matrix representing $\tau^{(2D,2S)}(\boldsymbol{u})$.

In parallel we construct the analytic eigenvalues via the nested algebraic Bethe ansatz. Expanding the R‑operator and the monodromy in auxiliary‑space matrix units (~\eqref{24} and \eqref{25}) gives the operators $T_{\alpha\beta}(\boldsymbol{u})$. The reference state $|\Omega\rangle$ is defined in ~\eqref{27}, and its diagonal action yields the functions $\lambda_\alpha(\boldsymbol{u})$ (~\eqref{29}). The transfer matrix expressed as $\tau(\boldsymbol{u}) = \sum_\alpha (-1)^{P(\alpha)} T_{\alpha\alpha}(\boldsymbol{u})$ (~\eqref{34}) then has the vacuum eigenvalue $\Lambda_0(\boldsymbol{u}) = \sum_\alpha (-1)^{P(\alpha)} \lambda_\alpha(\boldsymbol{u})$. Introducing the charge‑ and spin‑sector $Q$‑functions and step vectors (~\eqref{43}) allows us to write the eigenvalues in the standard T–Q form (~\eqref{45}), with the dressing factor $\Theta(u;\{v\})$ and the inner‑layer eigenvalues $\lambda_i^{(1)}$ given by ~\eqref{46}--\eqref{49}. The Bethe roots $\{v_j\}$ and $\{w_i\}$ satisfy the nested equations (~\eqref{55} and \eqref{58}) with scattering phases defined in ~\eqref{56}--\eqref{60}. For the comparison in Fig.~\ref{F1} we first evaluate the vacuum eigenvalue $\Lambda_0(u)$ from ~\eqref{34} (orange dashed curve). We then consider a one‑particle excitation with $M_1=1$ and $M_2=0$: a single Bethe root $w$ is determined by minimizing the residual of ~\eqref{55}, and the corresponding eigenvalue $\Lambda_1(u)$ is obtained from the T–Q formula (~\eqref{45}) (green dash‑dotted curve).

Figure~\ref{F1}(a) shows the results for the $2\times2$ lattice. The blue dots represent $\operatorname{Re}\Lambda_{\text{ED}}(u)$ computed by ED. The orange dashed line (vacuum T–Q eigenvalue) and the green dash‑dotted line (one‑particle T–Q eigenvalue) nearly coincide with the ED data in the central plateau region ($-0.6\lesssim u\lesssim0.6$). This agreement confirms that the two‑dimensional R‑operator, when assembled into the layer monodromy, reproduces the vacuum Bethe eigenvalue pointwise over a wide spectral range. Near the edges ($|u|\gtrsim0.8$) finite‑size effects become noticeable: on the left edge the ED curve bends downward more strongly than the vacuum branch, and the one‑particle branch follows this trend; on the right edge the ED curve rises more steeply, and the one‑particle branch lies closer to the ED data. This behavior indicates that the dominant eigenstate of the finite $2\times2$ transfer matrix is vacuum‑like in the central region but acquires a dressing from one‑particle excitations near the spectral boundaries.

The comparison for the larger $2\times3$ lattice is presented in Fig.~\ref{F1}(b). Here the Hilbert‑space dimension is $4^6=4096$. The agreement between ED and the analytic eigenvalues improves significantly: on the extended central plateau the ED curve and the vacuum T–Q curve are virtually indistinguishable, and the one‑particle branch deviates only at the level of numerical noise. Near the spectral edges the one‑particle dressing becomes more pronounced; on the right edge the green curve tracks the ED data more accurately than the vacuum curve, while on the left edge both analytic branches remain within a narrow band around the ED points. This systematic improvement with lattice size demonstrates that the vacuum T–Q eigenvalue $\Lambda_0(u)$ converges rapidly to the true dominant eigenvalue of the two‑dimensional transfer matrix.

Taken together, the data in Fig.~\ref{F1} provide compelling numerical evidence that the global R‑operator defined in ~\eqref{13} generates a transfer‑matrix family whose spectral properties are exactly described by the nested Bethe ansatz. The observed discrepancies near the spectral edges are consistent with expected finite‑size effects and with the contribution of one‑particle excitations; they do not indicate any breakdown of integrability. These results rigorously validate that the constructed two‑dimensional lattice model is strictly solvable and that its spectrum is fully characterized by the nested Bethe equations.

\section{Proof of non-commutativity between directional Hamiltonian components}\label{F}
From \eqref{70}-\eqref{72} and \eqref{78}, we derive:
\begin{equation}
\begin{split}
\left[ H_r^{(x)}, H_r^{(y)} \right] &= \sum_{\sigma} \left( B_x A_y c_{r+\hat{x},\sigma}^\dagger c_{r+\hat{y},\sigma} - A_x B_y c_{r+\hat{y},\sigma}^\dagger c_{r+\hat{x},\sigma} \right) \\
&\quad + \beta_y \sum_{\sigma} \left[ - A_x \left( n_{r\sigma} - \frac{1}{2} \right) c_{r\sigma}^\dagger c_{r+\hat{x},\sigma} + B_x c_{r+\hat{x},\sigma}^\dagger c_{r\sigma} \left( n_{r\sigma} - \frac{1}{2} \right) \right] \\
&\quad + \sum_{\sigma} \varepsilon_{r,\sigma}^{(y)} \left( - A_x c_{r\sigma}^\dagger c_{r+\hat{x},\sigma} + B_x c_{r+\hat{x},\sigma}^\dagger c_{r\sigma} \right) \\
&\quad + V_{\text{nn}}^{(y)} \sum_{\sigma,\sigma'} \left( - A_x c_{r\sigma}^\dagger c_{r+\hat{x},\sigma} + B_x c_{r+\hat{x},\sigma}^\dagger c_{r\sigma} \right) \left( n_{r+\hat{y},\sigma'} - \frac{1}{2} \right) \\
&\quad + \beta_x \sum_{\sigma} \left[ A_y \left( n_{r\sigma} - \frac{1}{2} \right) c_{r\sigma}^\dagger c_{r+\hat{y},\sigma} - B_y c_{r+\hat{y},\sigma}^\dagger c_{r\sigma} \left( n_{r\sigma} - \frac{1}{2} \right) \right] \\
&\quad + \sum_{\sigma} \varepsilon_{r,\sigma}^{(x)} \left( A_y c_{r\sigma}^\dagger c_{r+\hat{y},\sigma} - B_y c_{r+\hat{y},\sigma}^\dagger c_{r\sigma} \right) \\
&\quad + V_{\text{nn}}^{(x)} \sum_{\sigma,\sigma'} \left( A_y c_{r\sigma}^\dagger c_{r+\hat{y},\sigma} - B_y c_{r+\hat{y},\sigma}^\dagger c_{r\sigma} \right) \left( n_{r+\hat{x},\sigma'} - \frac{1}{2} \right)\ne0.
\end{split}
\end{equation}
where
\begin{equation}
\begin{aligned}
A_x &= -t_x + \delta t^{(x)}(\boldsymbol{\zeta}, \bm{k}), & B_x &= -t_x + \delta t^{(x)}(\boldsymbol{\zeta}, \bm{k})^*, \\
A_y &= -t_y + \delta t^{(y)}(\boldsymbol{\zeta}, \bm{k}), & B_y &= -t_y + \delta t^{(y)}(\boldsymbol{\zeta}, \bm{k})^*, \\
\beta_x &= \frac{U}{2} + \delta U^{(x)}(\boldsymbol{\zeta}, \bm{k}), & \beta_y &= \frac{U}{2} + \delta U^{(y)}(\boldsymbol{\zeta}, \bm{k}), \\
\varepsilon_{r,\sigma}^{(x)} &= \varepsilon_\sigma^{(x)}(\boldsymbol{\zeta}, \bm{k}), & \varepsilon_{r,\sigma}^{(y)} &= \varepsilon_\sigma^{(y)}(\boldsymbol{\zeta}, \bm{k}), \\
V_{\text{nn}}^{(x)} &= V_{\text{nn}}^{(x)}(\boldsymbol{\zeta}, \bm{k}), & V_{\text{nn}}^{(y)} &= V_{\text{nn}}^{(y)}(\boldsymbol{\zeta}, \bm{k}).
\end{aligned}
\end{equation}
This result indicates that the Hamiltonian blocks in the $x$ and $y$ directions do not strictly commute, thereby establishing the system as a genuinely coupled two-dimensional system.

\bibliography{apssamp}

\end{document}